\documentclass[12pt]{iopart}
\usepackage{iopams}
\newtheorem{theorem}{Theorem}
\newtheorem{lemma}{Lemma}
\newtheorem{remark}{Remark}

\begin{document}

\title[Torsion waves in metric--affine field theory]
{Torsion waves in metric--affine field theory}

\author{Alastair D King\dag\ and Dmitri Vassiliev\ddag}

\address{Department of Mathematical Sciences, University of Bath,
Bath BA2 7AY, UK}

\eads{\dag\mailto{A.D.King@maths.bath.ac.uk}}
\eads{\ddag\mailto{D.Vassiliev@bath.ac.uk}}

\begin{abstract}
The approach of metric--affine field theory is to
define spacetime
as a real oriented 4-manifold equipped with a metric
and an affine connection.
The 10 independent components of the metric tensor
and the 64 connection coefficients are the unknowns of the theory.
We write the Yang--Mills action for the affine connection
and vary it both with respect to
the metric and the connection.
We find a family of spacetimes which are
stationary points.
These spacetimes are waves of torsion in Minkowski space.
We then find a special subfamily of spacetimes with zero Ricci
curvature;
the latter condition is the Einstein
equation describing the absence of sources of gravitation.
A detailed examination
of this special subfamily suggests
the possibility of using it to model the neutrino.
Our model naturally contains only two distinct types of particles
which may be identified with
left-handed neutrinos and
right-handed antineutrinos.
\end{abstract}


\submitto{\CQG}


\section{Main results}
\label{intro}

We consider spacetime to be a real oriented 4-manifold $M$
equipped with a non-degenerate symmetric metric $g$
and an affine connection $\Gamma$.
The 10 independent
components of the metric tensor $g_{\mu\nu}$
and the 64 connection coefficients ${\Gamma^\lambda}_{\mu\nu}$
are the unknowns, as is the manifold $M$ itself.
This approach is known as metric--affine field theory.
Its origins lie in the works of authors such as \'E Cartan,
A S Eddington, A~Einstein, T Levi-Civita, E Schr\"odinger and H Weyl;
see, for example, Appendix II in \cite{einsteinthemeaning},
or \cite{schrodingerbook}.
Reviews of the more recent
work in this area can be found in
\cite{mielkebook,hehlreview,gronwald,hehlexact}.

The Yang--Mills action for the affine connection is
\begin{equation}
\label{yangmillsaction}
S_{\rm YM}:=
\int{R^\kappa}_{\lambda\mu\nu}\,
{R^\lambda}_\kappa{}^{\mu\nu}
\end{equation}
where $R$ is the Riemann curvature tensor
(\ref{riemanncurvaturetensor}).
Variation of (\ref{yangmillsaction}) with respect to the metric
$g$ and the connection $\Gamma$ produces Euler--Lagrange
equations which we, for the time being, will write symbolically as
\begin{equation}
\label{complementaryyangmillsequation}
\partial S_{\rm YM}/\partial g=0\,,
\end{equation}
\begin{equation}
\label{yangmillsequation}
\partial S_{\rm YM}/\partial\Gamma=0\,.
\end{equation}
Equation (\ref{yangmillsequation}) is the Yang--Mills equation
for the affine connection.
Equation (\ref{complementaryyangmillsequation}) does not have
an established name; we will call it the
\emph{complementary Yang--Mills equation}.

Our initial objective is the study of the combined system
(\ref{complementaryyangmillsequation}), (\ref{yangmillsequation}).
This is a system of 74 real non-linear partial differential equations
with 74 real unknowns.

In order to state our results we will require the Maxwell equation
\begin{equation}
\label{maxwellequation}
\delta\rmd u=0
\end{equation}
as well as the \emph{polarized Maxwell equation}
\begin{equation}
\label{polarizedmaxwellequation}
*\rmd u=\alpha\rmi\rmd u\,,
\end{equation}
$\alpha=\pm1$; here $u$ is the unknown vector function.
In calling (\ref{polarizedmaxwellequation})
the polarized Maxwell equation we are motivated by the fact
that any solution of
(\ref{polarizedmaxwellequation}) is a solution of
(\ref{maxwellequation}).
We call a solution $u$ of the Maxwell equation (\ref{maxwellequation})
non-trivial if $\rmd u\not\equiv0$.

If the metric is given and the connection is
known to be metric compatible
then the connection coefficients are uniquely determined by
torsion (\ref{torsiontensor}) or contortion (\ref{contortiontensor}).
The choice of torsion or contortion
for the purpose of describing a metric compatible connection
is purely
a matter of convenience as the two are expressed one via the
other in accordance with formulae (\ref{torsionandcontortion}).

We define Minkowski space ${\mathbb M}^4$
as a real 4-manifold with
a global coordinate system $(x^0,x^1,x^2,x^3)$ and metric
$g_{\mu\nu}=\rm{diag}(+1,-1,-1,-1)\,$.
Our definition of ${\mathbb M}^4$
specifies the manifold $M$ and the metric $g$,
but does not specify the connection $\Gamma$.

Our first result is

\begin{theorem}
\label{theorem1}
Let $u$ be a complex-valued vector function on ${\mathbb M}^4$ which is
a non-trivial plane wave solution of the polarized Maxwell equation
(\ref{polarizedmaxwellequation}),
let $L\ne0$ be a constant complex antisymmetric tensor satisfying
\begin{equation}
\label{polarizedtensor}
*L=\tilde\alpha\rmi L\,,
\end{equation}
$\tilde\alpha=\pm1$,
and let $\Gamma$ be the metric compatible connection
corresponding to contortion
\begin{equation}
\label{factorizedcontortion}
{K^\lambda}_{\mu\nu}
={\rm Re}(u_\mu{L^\lambda}_\nu)\,.
\end{equation}
Then the spacetime $\{{\mathbb M}^4,\Gamma\}$
is a solution of the system of equations
(\ref{complementaryyangmillsequation}), (\ref{yangmillsequation}).
\end{theorem}

\begin{remark}
In abstract Yang--Mills theory it is not customary to consider
the equation (\ref{complementaryyangmillsequation}) because there
is no guarantee that this would lead
to physically meaningful results.
As an illustration let us examine the Maxwell equation
(\ref{maxwellequation})
for real-valued vector functions on a Lorentzian manifold,
which is the simplest example of a Yang--Mills equation.
Straightforward calculations show that it
does not have non-trivial solutions which are stationary points
of the Maxwell action with respect to the variation of the metric.
\end{remark}

It is easy to see that the connections from Theorem \ref{theorem1} are not
flat, i.e., $R\not\equiv0$.

The non-trivial
plane wave solutions of (\ref{polarizedmaxwellequation})
can, of course, be written down explicitly: up to a proper
Lorentz transformation they are
\begin{equation}
\label{formulaforu}
u(x)\,=\,w\,\rme^{-\rmi k\cdot x}
\end{equation}
where
\begin{equation}
\label{formulaforwk}
w_\mu=C(0,1,-\alpha\rmi,0)\,,
\qquad
k_\mu=\beta(1,0,0,1)\,,
\qquad
\end{equation}
$\beta=\pm 1$, and $C$ is an arbitrary positive constant (amplitude).

Let us now introduce an additional equation into our model:
\begin{equation}
\label{einsteinequation}
Ric=0
\end{equation}
where $Ric$ is the Ricci curvature tensor.
This is the Einstein equation describing the absence of
sources of gravitation.

\begin{remark}
If the connection is that of Levi-Civita then
(\ref{einsteinequation}) implies (\ref{yangmillsequation}).
In the general case equations (\ref{yangmillsequation})
and (\ref{einsteinequation}) are independent.
\end{remark}

The question we are about to address
is whether there are any spacetimes
which simultaneously satisfy
the Yang--Mills equation (\ref{yangmillsequation}),
the complementary Yang--Mills equation
(\ref{complementaryyangmillsequation}),
and the Einstein equation (\ref{einsteinequation}).
More specifically, we are interested in spacetimes whose
connections are not flat and not Levi-Civita connections.

The following theorem provides an affirmative answer to the above
question.

\begin{theorem}
\label{theorem2}
A spacetime from Theorem \ref{theorem1} satisfies
equation (\ref{einsteinequation}) if and only if $L$
is proportional to $\left.(\rmd u)\right|_{x=0}$,
in which case torsion equals contortion
up to a natural reordering of indices:
\begin{equation}
\label{torsionequalscontortion}
T_{\lambda\mu\nu}=K_{\mu\lambda\nu}.
\end{equation}
\end{theorem}

When describing the spacetimes from Theorem \ref{theorem2}
it is convenient to take $L=\rmd u$
rather than $L=\left.(\rmd u)\right|_{x=0}\,$. This leads to a
rescaling of the wave vector $k$ which can, of course, be
incorporated into a Lorentz transformation. Thus,
the torsion of spacetimes from
Theorem \ref{theorem2} can be written as
\begin{equation}
\label{torsionexplicit}
T_{\lambda\mu\nu}
={\rm Re}(u_\lambda(\rmd u)_{\mu\nu})\,.
\end{equation}

The paper has the following structure.

Sections \ref{solvingyangmills} and \ref{solvingcomplementaryyangmills}
contain the proof of Theorem \ref{theorem1}, whereas Section
\ref{solvingeinstein} contains the proof of Theorem
\ref{theorem2}. The central elements of our construction
are the linearization ansatz (Lemma \ref{commutingcorollary})
and the double duality ansatz (Lemma \ref{complementarylemma}).

The rest of the paper is a detailed examination of the
spacetimes from Theorem \ref{theorem2}.

In Section \ref{invariant} we establish general invariant
properties of the spacetimes from Theorem \ref{theorem2}.
In particular,
it turns out (Lemma \ref{curvaturelemma})
that their Riemann curvature tensors possess
\emph{all} the symmetry properties of the ``usual''
curvature tensors generated by Levi-Civita connections.
This means that in observing such connections
we might be led to believe (mistakenly) that we live
in a Levi-Civita universe.

In Section \ref{weyl} we show that the Riemann curvature
tensors corresponding to space\-times from Theorem \ref{theorem2}
have an algebraic
structure which makes them equivalent to bispinors.
It turns out (Lemma \ref{weyllemma})
that these bispinors satisfy the Weyl equation, which suggests
the possibility of interpreting such spacetimes
as a model for the neutrino.
Our model naturally contains only two distinct types of particles
which may be identified with
left-handed neutrinos and
right-handed antineutrinos.

Finally, in Section \ref{doublestar} we compare our results
with those of Einstein \cite{einsteindoublestar} who
performed a double duality analysis of Riemann curvatures
with the aim of modelling elementary particles.
We show that the spacetimes from Theorem \ref{theorem2}
are in agreement with the results of Einstein's analysis,
in that we get the sign predicted by Einstein.

\section{Notation}
\label{notaion}

We denote $\partial_\mu=\partial/\partial x^\mu$ and define
the covariant derivative of
a vector function as
$\nabla_{\!\mu}v^\lambda:=\partial_\mu v^\lambda
+{\Gamma^\lambda}_{\mu\nu}v^\nu$.
We define the torsion tensor as
\begin{equation}
\label{torsiontensor}
{T^\lambda}_{\mu\nu}:=
{\Gamma^\lambda}_{\mu\nu}-{\Gamma^\lambda}_{\nu\mu}\,,
\end{equation}
the Riemann curvature tensor as
\begin{equation}
\label{riemanncurvaturetensor}
{R^\kappa}_{\lambda\mu\nu}:=
\partial_\mu{\Gamma^\kappa}_{\nu\lambda}
-\partial_\nu{\Gamma^\kappa}_{\mu\lambda}
+{\Gamma^\kappa}_{\mu\eta}{\Gamma^\eta}_{\nu\lambda}
-{\Gamma^\kappa}_{\nu\eta}{\Gamma^\eta}_{\mu\lambda}\,,
\end{equation}
and the Ricci curvature tensor as
$Ric_{\lambda\nu}:={R^\kappa}_{\lambda\kappa\nu}$.

We employ the usual convention of raising or lowering tensor indices
by contraction with the contravariant or covariant metric
tensor. Some care is, however, required when
performing covariant differentiation:
the operations of raising or lowering of indices do not
commute with the operation of covariant differentiation
unless the connection is metric compatible.

By $\rmd$ we denote the exterior derivative
and by $\delta$ its adjoint.
Of course, these operators do not depend on the
connection.

Given a scalar function $f$
we write for brevity
\begin{equation}
\label{integralofascalar}
\int f:=\int_Mf\,\sqrt{|\det g|}
\,dx^0dx^1dx^2dx^3\,,
\quad
\det g:=\det(g_{\mu\nu})\,.
\end{equation}

Throughout the paper we work only in coordinate systems with
positive orientation. Moreover, when we restrict our consideration
to Minkowski space we assume that our coordinate frame is obtained
from a given reference frame by a proper Lorentz transformation.
We use these conventions when defining
the notions of left-handedness and right-handedness,
as well as those of the forward and backward light cone.

We define the Hodge star as
$(*Q)_{\mu_{q+1}\ldots\mu_4}\!:=(q!)^{-1}\,\sqrt{|\det g|}\,
Q^{\mu_1\ldots\mu_q}\varepsilon_{\mu_1\ldots\mu_4}$
where $\varepsilon$ is the totally antisymmetric quantity,
$\varepsilon_{0123}:=+1$.

When dealing with a
connection which is compatible with
a given metric it is convenient to introduce
the \emph{contortion} tensor
\begin{equation}
\label{contortiontensor}
{K^\lambda}_{\mu\nu}:=
{\Gamma^\lambda}_{\mu\nu}-
\left\{{{\lambda}\atop{\mu\nu}}\right\}
\end{equation}
where
$\left\{{{\lambda}\atop{\mu\nu}}\right\}:=
\frac12g^{\lambda\kappa}
(\partial_\mu g_{\nu\kappa}
+\partial_\nu g_{\mu\kappa}
-\partial_\kappa g_{\mu\nu})$
is the Christoffel symbol.
Contortion has the antisymmetry property
$K_{\lambda\mu\nu}=-K_{\nu\mu\lambda}\,.$
A metric and contortion uniquely determine the
metric compatible connection.
Torsion and contortion are related as
\begin{equation}
\label{torsionandcontortion}
{T^\lambda}_{\mu\nu}=
{K^\lambda}_{\mu\nu}-{K^\lambda}_{\nu\mu}\,,
\qquad
{K^\lambda}_{\mu\nu}=\case12
\bigl(
T{}^\lambda{}_{\mu\nu}+T{}_\mu{}^\lambda{}_\nu+T{}_\nu{}^\lambda{}_\mu
\bigr)\,,
\end{equation}
see formula (7.35) in \cite{nakahara}.

The remainder of this section is devoted to the special
case of Minkowski space.

Lorentz transformations
are assumed to be ``passive'' in the sense
that we transform the coordinate system and not the tensors or
spinors themselves.

Consider a complex-valued tensor or spinor
function of the form
$\,{\rm const}\times\rme^{-\rmi k\cdot x}\,$ where
$k\ne0$ is a constant real vector and $k\cdot x:=k_\mu x^\mu$.
We call such a function a plane wave and
the vector $k$ a wave vector.
In defining a plane wave as
$\sim\rme^{-\rmi k\cdot x}$ rather than $\sim\rme^{\rmi k\cdot x}$
we follow the convention of \cite{LL2,LL4,itzykson}.
We say that a lightlike wave vector $k$ lies on the
forward (respectively, backward) light cone if $k_0>0$
(respectively, $k_0<0$).

A bispinor is a column of four complex numbers
$(\begin{array}{cccc}
\xi^1&\xi^2&\eta_{\dot1}&\eta_{\dot2}
\end{array})^T$
which change under Lorentz transformations
in a particular way, see Section 18 in \cite{LL4}
for details.
The Pauli and Dirac matrices are
\[
\fl
\sigma^0=\left(\begin{array}{cc}1&0\\0&1\end{array}\right),\qquad
\sigma^1=\left(\begin{array}{cc}0&1\\1&0\end{array}\right),\qquad
\sigma^2=\left(\begin{array}{cc}0&-i\\ i&0\end{array}\right),\qquad
\sigma^3=\left(\begin{array}{cc}1&0\\0&-1\end{array}\right),
\]
\[
\gamma^0=\left(\begin{array}{cc}0&-\sigma^0\\-\sigma^0&0
\end{array}\right),\qquad
\gamma^j=\left(\begin{array}{cc}0&\sigma^j\\-\sigma^j&0
\end{array}\right),\quad j=1,2,3,
\]
\[
\gamma^5=i\gamma^0\gamma^1\gamma^2\gamma^3
=\left(\begin{array}{cc}\sigma^0&0\\0&-\sigma^0\end{array}\right).
\]
We chose the sign of $\gamma^5$ as in
\cite{itzykson} (in \cite{LL4} it is opposite).

\section{Solving the Yang--Mills equation}
\label{solvingyangmills}

When dealing with the Yang--Mills equation it is
convenient to use matrix notation to hide two indices:
$R_{\mu\nu}={R^\kappa}_{\lambda\mu\nu}$,
$\Gamma_\rho={\Gamma^\kappa}_{\rho\lambda}$,
with $\kappa$ enumerating the rows and $\lambda$ the columns.
Formulae (\ref{yangmillsaction}), (\ref{riemanncurvaturetensor})
can be rewritten in this notation as
\begin{equation}
\label{yangmillsaction1}
S_{\rm YM}:=
\int\tr(R_{\mu\nu}\,R^{\mu\nu})\,,
\end{equation}
\begin{equation}
\label{riemanncurvaturetensor1}
R_{\mu\nu}=\partial_\mu\Gamma_\nu-\partial_\nu\Gamma_\mu
+[\Gamma_\mu,\Gamma_\nu]\,,
\end{equation}
where
$\tr L:={L^\kappa}_\kappa$ (trace of a matrix)
and
${[L,N]^\tau}_\lambda:=
{L^\tau}_{\kappa}{N^\kappa}_\lambda
-{N^\tau}_{\kappa}{L^\kappa}_\lambda$
(commutator of matrices).
Straightforward analysis of formulae
(\ref{yangmillsaction1}), (\ref{riemanncurvaturetensor1}),
(\ref{integralofascalar})
shows that the Yang--Mills equation
which we initially wrote down symbolically as
(\ref{yangmillsequation})
is actually
\begin{equation}
\label{yangmillsequation1}
(\partial_\mu+[\Gamma_\mu,\,\cdot\,])
(\sqrt{|\det g|}\,R^{\mu\nu})=0\,.
\end{equation}

From now on we work
only in Minkowski space and only with metric compatible connections.
This leads to a number of simplifications.
Connection coefficients now coincide with contortion,
for which we continue using matrix notation
$K_\rho={K^\kappa}_{\rho\lambda}$. Formula
(\ref{riemanncurvaturetensor1}) becomes
\begin{equation}
\label{riemanncurvaturetensor2}
R_{\mu\nu}=\partial_\mu K_\nu-\partial_\nu K_\mu+[K_\mu,K_\nu]\,,
\end{equation}
and the Yang--Mills equation
(\ref{yangmillsequation1}) becomes
\begin{equation}
\label{yangmillsequation2}
(\partial_\nu+[K_\nu,\,\cdot\,])R^{\mu\nu}=0\,.
\end{equation}

The Yang--Mills equation (\ref{yangmillsequation2}) appears to be
overdetermined as it is a system of 64 equations with only 24 unknowns
(24 is the number of independent components of the contortion
tensor).
However 40 of the 64 equations
are automatically fulfilled. This is a consequence of the fact
that the 6-dimensional
Lie algebra of real antisymmetric rank 2 tensors is
a subalgebra of the 16-dimensional
general Lie algebra of real rank 2 tensors.

The fundamental difficulty with the Yang--Mills equation
is that it is nonlinear
with respect to the unknown contortion $K$.
The following lemma
plays a crucial role in our construction by
allowing us to get rid of the nonlinearities.

\begin{lemma}
\label{commutinglemma}
If $L$ is an eigenvector of the Hodge star
then $[{\rm Re}L,{\rm Im}L]=0$.
\end{lemma}

\noindent
{\it Proof of Lemma \ref{commutinglemma}\ }
The result follows from the general formula $[*L,N]=*[L,N]$.
$\square$

Lemma \ref{commutinglemma} can be rephrased in the following way:
the 6-dimensional Lie algebra of
real anti\-symmetric rank 2 tensors has 2-dimensional abelian
subalgebras which can be explicitly described in terms of
the eigenvectors of the Hodge star.

Lemma \ref{commutinglemma} immediately implies
the following \emph{linearization ansatz.}

\begin{lemma}
\label{commutingcorollary}
Suppose contortion is of the form (\ref{factorizedcontortion})
where $u$ is a complex-valued vector function
and $L\ne0$ is a constant complex antisymmetric
tensor satisfying (\ref{polarizedtensor}).
Then the nonlinear terms in
the formula for Riemann curvature (\ref{riemanncurvaturetensor2})
and in the Yang--Mills equation
(\ref{yangmillsequation2}) vanish.
\end{lemma}

Substituting
(\ref{factorizedcontortion}) into (\ref{riemanncurvaturetensor2})
and the latter into (\ref{yangmillsequation2}) we see that
the Yang--Mills equation reduces to the Maxwell equation
(\ref{maxwellequation}) for the complex-valued vector function $u$.

\section{Solving the complementary Yang--Mills equation}
\label{solvingcomplementaryyangmills}

Straightforward analysis of formulae
(\ref{yangmillsaction}), (\ref{integralofascalar})
shows that the complementary Yang--Mills equation
which we initially wrote down symbolically as
(\ref{complementaryyangmillsequation})
is actually
\begin{equation}
\label{complementaryyangmillsequation1}
H-\case14\,(\tr H)\,\delta=0
\end{equation}
where
$\,H=
{H_\nu}^\rho:={R^\kappa}_{\lambda\mu\nu}{R^\lambda}_\kappa{}^{\mu\rho}\,$
and $\,\delta={\delta_\nu}^\rho\,$ is the identity tensor.
Note the important difference between the Yang--Mills equation
(\ref{yangmillsequation1}) and
the complementary Yang--Mills equation
(\ref{complementaryyangmillsequation1}):
equation (\ref{yangmillsequation1}) is linear in curvature,
whereas (\ref{complementaryyangmillsequation1}) is quadratic.

Equation (\ref{complementaryyangmillsequation1}) was
written down without any assumptions on the connection.
We, however, will be interested in solving
(\ref{complementaryyangmillsequation1}) in the class of
spacetimes with metric compatible connections, in which case
the Riemann curvature tensor has the symmetries
\begin{equation}
\label{complementary1}
R_{\kappa\lambda\mu\nu}
=-R_{\lambda\kappa\mu\nu}
=-R_{\kappa\lambda\nu\mu}\,.
\end{equation}

Let $\mathcal{R}$ be the 36-dimensional linear
space  of real rank 4 tensors satisfying
(\ref{complementary1}).
We define in $\mathcal{R}$ the following two commuting
endomorphisms
\[
R\to{}^*\!R\,,
\qquad
({}^*\!R)_{\kappa\lambda\mu\nu}:=\case12\,\sqrt{|\det g|}
\ \varepsilon^{\kappa'\lambda'}{}_{\kappa\lambda}\,
R_{\kappa'\lambda'\mu\nu}\,,
\]
\[
R\to R^*\,,
\qquad
(R^*)_{\kappa\lambda\mu\nu}:=\case12\,\sqrt{|\det g|}
\ R_{\kappa\lambda\mu'\nu'}\,
\varepsilon^{\mu'\nu'}{}_{\mu\nu}\,,
\]
and we also consider their composition
\begin{equation}
\label{complementary2}
R\to{}^*\!R^*\,.
\end{equation}
Clearly, the endomorphism (\ref{complementary2})
has eigenvalues $\pm1$.

\begin{remark}
It is easy to see that
the endomorphism (\ref{complementary2}) is well defined even
if the manifold is not orientable.
This observation is related to
a much deeper fact established in \cite{lanczos}:
the rank 8 tensor
$\,(\det g)\,\varepsilon_{\kappa'\lambda'\kappa\lambda}
\,\varepsilon_{\mu'\nu'\mu\nu}\,$
is a purely metrical quantity, i.e., it is expressed via the
metric tensor.
\end{remark}

The following lemma is the \emph{double duality ansatz}
which reduces the complementary Yang--Mills equation to
an equation linear in curvature.

\begin{lemma}
\label{complementarylemma}
If $R\in\mathcal{R}$ is an eigenvector of (\ref{complementary2})
then it satisfies
(\ref{complementaryyangmillsequation1}).
\end{lemma}

\noindent
{\it Proof of Lemma \ref{complementarylemma}\ }
We have
\begin{equation}
\label{complementary3}
\fl
{H_\nu}^\rho
={R^\kappa}_{\lambda\mu\nu}{R^\lambda}_\kappa{}^{\mu\rho}
=R_{\kappa\lambda\nu\mu}R^{\kappa\lambda\mu\rho}
=\case12
(R_{\kappa\lambda\nu\mu}R^{\kappa\lambda\mu\rho}
+({}^*\!R^*)_{\kappa\lambda\nu\mu}({}^*\!R^*)^{\kappa\lambda\mu\rho})
\,.
\end{equation}
For antisymmetric rank 2 tensors we have the identities
\[
(*L)_{\kappa\lambda}(*N)^{\kappa\lambda}
=-L_{\kappa\lambda}N^{\kappa\lambda}\,,
\]
\[
(*L)_{\nu\mu}(*N)^{\mu\rho}+(*N)_{\nu\mu}(*L)^{\mu\rho}
=L_{\nu\mu}N^{\mu\rho}+N_{\nu\mu}L^{\mu\rho}
+L_{\mu\tau}N^{\mu\tau}{\delta_\nu}^\rho\,,
\]
so formula (\ref{complementary3}) can be continued as
\begin{eqnarray*}
\fl
{H_\nu}^\rho
=\case12
(R_{\kappa\lambda\nu\mu}R^{\kappa\lambda\mu\rho}
-(R^*)_{\kappa\lambda\nu\mu}(R^*)^{\kappa\lambda\mu\rho})
\\
=\case14
(
(R_{\kappa\lambda\nu\mu}R^{\kappa\lambda\mu\rho}+
{R^{\kappa\lambda}}_{\nu\mu}{R_{\kappa\lambda}}^{\mu\rho})
-
((R^*)_{\kappa\lambda\nu\mu}(R^*)^{\kappa\lambda\mu\rho}
+{(R^*)^{\kappa\lambda}}_{\nu\mu}{(R^*)_{\kappa\lambda}}^{\mu\rho})
)
\\
=-\case14R_{\kappa\lambda\mu\tau}R^{\kappa\lambda\mu\tau}
{\delta_\nu}^\rho\,.
\end{eqnarray*}
The tensor ${H_\nu}^\rho$ is proportional to the identity
tensor ${\delta_\nu}^\rho$, therefore it satisfies
(\ref{complementaryyangmillsequation1}).
$\square$

Let us now apply Lemma \ref{complementarylemma} to the spacetimes
constructed in the previous section. In view of Lemma
\ref{commutingcorollary} the Riemann curvature in this case is
\begin{equation}
\label{complementary4}
R_{\kappa\lambda\mu\nu}={\rm Re}
(L_{\kappa\lambda}(\rmd u)_{\mu\nu})
\end{equation}
where $L$ is an eigenvector of the Hodge star and $u$ is a
non-trivial
complex-valued solution of the Maxwell equation (\ref{maxwellequation}).
Clearly, (\ref{complementary4}) is an eigenvector of the
endomorphism (\ref{complementary2}) if and only if
$\rmd u$ is an eigenvector of the Hodge star. The latter means that
$u$ is a solution of the polarized Maxwell equation
(\ref{polarizedmaxwellequation}).
The proof of Theorem \ref{theorem1} is complete.

\section{Solving the Einstein equation}
\label{solvingeinstein}

Substituting (\ref{complementary4}) into the Einstein equation
(\ref{einsteinequation}) we get
${\rm Re}({L^\kappa}_\lambda(\rmd u)_{\kappa\nu})=0\,$.
As the expression under the $\,{\rm Re}\,$ sign is a plane wave,
the latter is equivalent to
\begin{equation}
\label{einstein1}
{L^\kappa}_\lambda(\left.(\rmd u)\right|_{x=0})_{\kappa\nu}=0\,.
\end{equation}
It is convenient to perform further calculations in the coordinate
system in which $u$ has the canonical form
(\ref{formulaforu}), (\ref{formulaforwk}).
Then
\begin{equation}
\label{einstein2}
(\rmd u)_{\kappa\nu}=C\,\rmi
\left(\begin{array}{cccc}
0&-1&\alpha\rmi&0\\
1&0&0&1\\
-\alpha\rmi&0&0&-\alpha\rmi\\
0&-1&\alpha\rmi&0
\end{array}\right)
\rme^{-\rmi\beta(x^0+x^3)}
\end{equation}
and (\ref{einstein1}) becomes an explicit system of linear
algebraic equations with respect to the unknown components
of the tensor $L$\,; namely, it is a system of 16 equations with
3 unknowns (recall that $L$ has to be an eigenvector
of the Hodge star). Elementary analysis
shows that equation (\ref{einstein1})
is satisfied if and only if $L$
is proportional to $\left.(\rmd u)\right|_{x=0}$.
Finally, formula (\ref{torsionequalscontortion})
is established by straightforward calculations
(see also Lemma \ref{torsioncontortionlemma} in the next section).
The proof of Theorem \ref{theorem2} is complete.

\section{Invariant properties of our solutions}
\label{invariant}

It is known \cite{hehlreview,gronwald,hehlexact}
that the 24-dimensional space of real torsions decomposes
into the following 3 irreducible subspaces:
tensor torsions, trace torsions, and axial torsions.
The dimensions of these subspaces
are 16, 4, and 4, respectively.

\begin{lemma}
\label{torsionlemma}
The torsions of spacetimes
from Theorem \ref{theorem2} are purely tensor.
\end{lemma}

\noindent
{\it Proof of Lemma \ref{torsionlemma}\ }
The trace component of a torsion tensor $T_{\lambda\mu\nu}$
is zero if ${T^\lambda}_{\lambda\nu}=0$,
and the axial component is zero if
$T_{\lambda\mu\nu}\,\varepsilon^{\lambda\mu\nu\kappa}=0$.
These identities are established by direct examination of
the explicit formulae (\ref{torsionexplicit}),
(\ref{formulaforu}), (\ref{formulaforwk}).
$\square$

It has been suggested \cite{vassiliev} to interpret
the axial component of torsion as the Hodge dual of
the electromagnetic
vector potential. If one takes this point of view
then Lemma~\ref{torsionlemma} implies that in spacetimes
from Theorem \ref{theorem2} the electromagnetic field
is zero.

Let us mention (without proof) the following useful general result.

\begin{lemma}
\label{torsioncontortionlemma}
Equation (\ref{torsionequalscontortion}) is satisfied
if and only if the axial component of torsion is zero.
\end{lemma}

Lemmas \ref{torsionlemma} and \ref{torsioncontortionlemma} imply that when
working with spacetimes from Theorem \ref{theorem2}
one can switch from contortion to torsion and back without acquiring
cumbersome expressions.

\begin{lemma}
\label{curvaturelemma}
The Riemann curvatures of spacetimes
from Theorem \ref{theorem2}
have all the sym\-metry properties
of Riemann curvatures in the Levi-Civita setting, that is,
\begin{equation}
\label{invariant1}
R_{\kappa\lambda\mu\nu}
=-R_{\lambda\kappa\mu\nu}
=-R_{\kappa\lambda\nu\mu}=R_{\mu\nu\kappa\lambda}\,,
\end{equation}
\begin{equation}
\label{invariant2}
R_{\kappa\lambda\mu\nu}
\,\varepsilon^{\kappa\lambda\mu\nu}=0\,.
\end{equation}
\end{lemma}

\noindent
{\it Proof of Lemma \ref{curvaturelemma}\ }
Let us define the complex Riemann curvature tensor
\begin{equation}
\label{invariant3}
{\mathbb C}\!R_{\kappa\lambda\mu\nu}:=
F_{\kappa\lambda}F_{\mu\nu}
\end{equation}
where
\begin{equation}
\label{invariant3F}
F:=\rmd u
\end{equation}
and $u$ is a plane wave solution of
(\ref{polarizedmaxwellequation}). Then the Riemann curvature
generated by torsion (\ref{torsionexplicit}) can be written as
\begin{equation}
\label{invariant4}
R={\rm Re}\,{\mathbb C}\!R
\end{equation}
(cf. (\ref{complementary4})).
Direct examination of formulae
(\ref{invariant3})--(\ref{invariant4}),
(\ref{einstein2})
establishes the
identities (\ref{invariant1}), (\ref{invariant2}).
$\square$

\section{Weyl's equation}
\label{weyl}

The torsions (and, therefore, spacetimes)
from Theorem \ref{theorem2} are described, up to
a proper Lorentz transformation and a scaling factor
$C>0$, by a pair of indices $\alpha,\beta=\pm1$;
see (\ref{torsionexplicit}), (\ref{formulaforu}), (\ref{formulaforwk}).
It may seem that this gives us 4 essentially different spacetimes.
However, formula (\ref{torsionexplicit}) contains the operation
of taking the real part and, as a result, the transformation
$\{\alpha,\beta\}\to\{-\alpha,-\beta\}$ does not change our torsion.
Thus, Theorem~\ref{theorem2} provides us with only two essentially
different spacetimes labelled by the product
$\alpha\beta=\pm1$. The purpose of this section
is to show that it is natural to interpret these two spacetimes
as the neutrino and antineutrino.

We base our interpretation on the analysis of the Riemann curvature
tensor.
We choose to deal with curvature rather than
with torsion because curvature
is an accepted physical observable.

We will work with
the complex curvature (\ref{invariant3}) rather than
the real curvature (\ref{invariant4}) because the complex
one has a simpler structure. Indeed, according
to formula (\ref{invariant3}) the complex Riemann curvature
tensor ${\mathbb C}\!R$ factorizes as the square of a rank 2
tensor $F$ and is, therefore, completely determined by it.

Working with the rank 2 tensor $F$ is much easier than with the
original rank 4 tensor ${\mathbb C}\!R$, but one would like
to simplify the analysis even further by factorizing $F$ itself.
It is impossible to factorize $F$ as the square of a vector
but it is possible to factorize $F$ as the square of a bispinor.

\begin{lemma}
\label{tensorbispinor}
A complex rank 2 antisymmetric tensor $F$ satisfying
the conditions
\begin{equation}
\label{weyl1}
F_{\mu\nu}F^{\mu\nu}=0,\qquad(*F)_{\mu\nu}F^{\mu\nu}=0
\end{equation}
is equivalent to a bispinor $\psi$,
the relationship between the two being
\begin{equation}
\label{weyl2}
F^{\mu\nu}=-\case\rmi 4
\,\psi^T\gamma^0\gamma^2\gamma^\mu\gamma^\nu\psi\,.
\end{equation}
\end{lemma}

\noindent
{\it Proof of Lemma \ref{tensorbispinor}\ }
Formula (\ref{weyl2}) is a special case of the general
equivalence relation
between rank 2 antisymmetric tensors
and rank 2 symmetric bispinors, see end of Section 19 in \cite{LL4}.
Conditions (\ref{weyl1}) are necessary and sufficient
for the factorization of the symmetric rank 2 spinors
as squares of rank 1 spinors.
$\square$

\begin{remark}
The corresponding text in the end of Section 19 in \cite{LL4}
contains mistakes. These can be corrected
by replacing everywhere $\rmi$ by $-\rmi.$
\end{remark}

\begin{remark}
For a given $F$ formula (\ref{weyl2}) defines
the individual spinors
$\xi=(\begin{array}{cc}
\xi^1&\xi^2
\end{array})^T$
and
$\eta=(\begin{array}{cc}
\eta_{\dot1}&\eta_{\dot2}
\end{array})^T$
uniquely up to choice of sign.
This is in agreement
with the general fact that a spinor does not have
a specific sign, see the beginning of Section 19 in \cite{LL4}.
\end{remark}

\begin{remark}
Conditions (\ref{weyl1}) are equivalent to
$\,\det F=0$, $\det*F=0$.
\end{remark}

Our particular tensor $F$ defined by
formula (\ref{invariant3F}) satisfies conditions (\ref{weyl1}).
Indeed, $F_{\mu\nu}F^{\mu\nu}=0$
is the statement that the complex scalar curvature is zero
(consequence of the complex Ricci curvature being zero),
whereas $(*F)_{\mu\nu}F^{\mu\nu}=0$
is the statement that the complex Riemann curvature
satisfies the cyclic sum identity, cf. (\ref{invariant2}).
Thus, the complex Riemann curvature tensor (\ref{invariant3})
has an algebraic structure which makes it equivalent to
a bispinor. We will now establish which equations
this bispinor satisfies.

We say that two solutions $u$ and $u'$
of the Maxwell equation (\ref{maxwellequation})
belong to the same equivalence class if $\rmd u=\rmd u'$.
We say that two bispinor functions $\psi$ and $\psi'$ belong
to the same equivalence class if $\psi=\pm\psi'$.

\begin{lemma}
\label{weyllemma}
Formula (\ref{weyl2}) establishes a one--to--one correspondence
between the equi\-valence classes of
non-trivial plane wave solutions of the
polarized Maxwell equation (\ref{polarizedmaxwellequation})
and of the system
\begin{equation}
\label{weyl3}
\gamma^\mu\partial_\mu\psi=0\,,
\end{equation}
\begin{equation}
\label{weyl4}
\gamma^5\psi=\alpha\psi\,.
\end{equation}
\end{lemma}

Equation (\ref{weyl3}) is, of course, the Weyl equation
(Dirac equation for massless particle).

\noindent
{\it Proof of Lemma \ref{weyllemma}\ }
If $u$ is a non-trivial plane wave solution of the
polarized Maxwell equation (\ref{polarizedmaxwellequation})
then, up to a proper Lorentz transformation,
our tensor $F$ is given by formula (\ref{einstein2}),
where $C$ is a positive constant.
If $\psi$ is a non-trivial plane wave solution of the system
(\ref{weyl3}), (\ref{weyl4})
then, up to a proper Lorentz transformation,
\begin{equation}
\label{weyl6}
\psi=\pm\sqrt{C\,\rmi}
\left(\begin{array}{c}
0\\1+\alpha\\\rmi-\alpha\rmi\\0
\end{array}\right)
\rme^{-\frac\rmi 2\beta(x^0+x^3)}
\end{equation}
where $C$ is a positive constant.
Straightforward calculations show that the tensor function
(\ref{einstein2}) and the bispinor function (\ref{weyl6})
are related in accordance with formula (\ref{weyl2}).
$\square$

The parameter $\beta=\pm1$ in formula (\ref{weyl6})
determines whether the wave vector
lies on the forward ($\beta=+1$)
or backward ($\beta=-1$) light cone.
Non-trivial plane wave solutions
of (\ref{weyl3}), (\ref{weyl4})
whose wave vector lies on the forward light cone
are called neutrinos
whereas those
whose wave vector lies on the backward light cone
are called antineutrinos.

The parameter $\alpha=\pm1$ in formula (\ref{weyl6}) determines
whether the solution is left- or right-handed. A neutrino is
said to be left-handed if $\alpha=-1$ and right-handed if
$\alpha=+1$. An antineutrino is said to be  left-handed if
$\alpha=+1$ and right-handed if $\alpha=-1$.

\begin{remark}
The above definitions of left- and right-handedness
are given in terms of helicity. See Section 2-4-3 in
\cite{itzykson} for a detailed explanation of
why one should use helicity rather than chirality for
these purposes.
\end{remark}

As explained in the beginning of this section,
the transformation
$\{\alpha,\beta\}\to\{-\alpha,-\beta\}$ does not change the
resulting spacetime.  This means
that the torsion wave which models
the left-handed neutrino is identical to that
for the left-handed antineutrino,
and the torsion wave which models
the right-handed neutrino is identical to that
for the right-handed antineutrino.
Thus, our model contains as many distinct types of
neutrinos as are currently observed experimentally.

\section{Einstein's double duality analysis}
\label{doublestar}

Let us examine in more detail the linear space
of Riemann curvatures ${\mathcal R}$ introduced in
Section \ref{solvingcomplementaryyangmills}.
For $R\in{\mathcal R}$ we define its transpose $R^T$
as $(R^T)_{\kappa\lambda\mu\nu}:=R_{\mu\nu\kappa\lambda}$.
We consider the following two
commuting endomorphisms in ${\mathcal R}$:
\begin{equation}
\label{doublestar2}
R\to R^T
\end{equation}
and (\ref{complementary2}).
The endomorphisms (\ref{doublestar2}) and (\ref{complementary2})
have no associated eigenvectors and
their eigenvalues are $\pm1$.
Therefore, ${\mathcal R}$
decomposes into a direct sum of 4 invariant subspaces
\begin{equation}
\label{doublestar4}
{\mathcal R}=
{\oplus}_{a,b=\pm}\ {\mathcal R}_{ab}\,,\qquad
{\mathcal R}_{ab}:=
\{R\in{\mathcal R}\,|\ R^T=aR,\ {}^*\!R{}^*=bR\}\,.
\end{equation}

The decomposition (\ref{doublestar4})
was suggested in \cite{rainich} and analyzed in
\cite{einsteindoublestar,lanczos}. Actually,
\cite{rainich,einsteindoublestar,lanczos}
dealt only with the case of a Levi-Civita connection,
but the generalization to an arbitrary metric compatible connection
is straightforward.
Lanczos called tensors $R\in{\mathcal R}$
self-dual (respectively, antidual) if ${}^*\!R{}^*=-R$
(respectively, ${}^*\!R{}^*=R$). Such a choice of terminology
is due to the fact that Einstein and Lanczos defined
their double duality endomorphism as
\begin{equation}
\label{doublestar5}
R\to({\rm sgn}\det g)\ {}^*\!R{}^*
\end{equation}
rather than as (\ref{complementary2}).
The advantage of (\ref{doublestar5}) is that this linear operator
is expressed via the metric tensor as a rational function.
The endomorphism (\ref{doublestar5}) is,
in a sense, even more invariant than (\ref{complementary2})
because it does not ``feel'' the signature of the metric.

\begin{lemma}
\label{rainichslemma}
{\rm (Rainich \cite{rainich})}
The subspaces ${\mathcal R}_{++}$ and ${\mathcal R}_{+-}$
have dimensions 9 and 12, respectively.
\end{lemma}

\begin{remark}
In Rainich's article the dimensions are actually given as 9 and 11.
The reason behind this is that Rainich imposed
on curvatures the cyclic sum condition
(\ref{invariant2}). This excludes from ${\mathcal R}_{+-}$
curvatures of the type
$R_{\kappa\lambda\mu\nu}={\rm const}\times
\varepsilon_{\kappa\lambda\mu\nu}$
and, therefore, reduces the dimension by 1.
\end{remark}

\begin{lemma}
\label{doublestarlemma}
{\rm (Einstein \cite{einsteindoublestar})}
Let $R\in{\mathcal R}_{++}$. Then the corresponding Ricci tensor
is symmetric and trace free.
Moreover, $R$ is uniquely determined by
its Ricci tensor and the metric tensor
according to the formula
\begin{equation}
\label{doublestar6}
R_{\kappa\lambda\mu\nu}=\case12
(g_{\kappa\mu}Ric_{\lambda\nu}+g_{\lambda\nu}Ric_{\kappa\mu}
-g_{\kappa\nu}Ric_{\lambda\mu}-g_{\lambda\mu}Ric_{\kappa\nu})\,.
\end{equation}
\end{lemma}

Einstein's goal in \cite{einsteindoublestar} was to
construct a relativistic  model for the electron;
note that this paper was published in 1927,
a year before Dirac published his equation.
(For a basic exposition of  \cite{einsteindoublestar} in English
see the Introduction in \cite{lanczos}.)
Einstein based his search for a mathematical
model on the decomposition
(\ref{doublestar4}). As in this particular paper Einstein
restricted his analysis to the case of a Levi-Civita connection he
had to make the choice between the invariant subspaces
${\mathcal R}_{++}$ and ${\mathcal R}_{+-}$.
The difference between these two invariant subspaces is fundamental:
it has nothing to do with
the choice between forward and backward light cones
or the choice of orientation of the coordinate system,
and, as a consequence,
it has nothing to do with the
notions of ``particle'' and ``antiparticle''
or the notions of ``left-handedness'' and ``right-handedness''.

Lemmas \ref{rainichslemma} and \ref{doublestarlemma}
led Einstein to the conclusion that
curvatures from ${\mathcal R}_{++}$ are too trivial
and the dimension of the subspace too low
to associate it with the electron.
Namely,
the main argument against ${\mathcal R}_{++}$ is that dimension 9 is not
enough to account for the 10
independent components of the energy--momentum
tensor.
This suggests that
if the electron were to be modelled in terms of General Relativity
then one would expect its
Riemann curvature to lie in the
invariant subspace ${\mathcal R}_{+-}$, that is,
satisfy the equation
\begin{equation}
\label{doublestar7}
{}^*\!R{}^*=-R\,.
\end{equation}

Formulae (\ref{invariant3})--(\ref{invariant4}),
(\ref{polarizedmaxwellequation}) imply that
the spacetimes from Theorem \ref{theorem2}
satisfy equation (\ref{doublestar7}).
Our paper falls short of constructing a metric--affine field
model for the electron, but,
nevertheless, we find it encouraging that
our metric--affine field model for the neutrino agrees with
the results of Einstein's analysis.

\ack

The authors are indebted to D~V Alekseevsky and
F~E Burstall for stimulating discussions.
The research of D Vassiliev was supported by a
Leverhulme Fellowship.

\end{document}